\documentclass[twocolumn,prl,showpacs,showkeys,amsmath,amssymb]{revtex4}
\usepackage{graphicx,amsmath,epsfig,latexsym,color}
\newcommand{\be}{\begin{equation}}
\newcommand{\ee}{\end{equation}}
\begin{document}
\title{Dispersion Compensation for Atom Interferometry}
\author{Tony D.\ Roberts, Alexander D.\ Cronin$^1$, Martin V.\ Tiberg,
David E.\ Pritchard} \affiliation{Massachusetts Institute of
Technology, Cambridge, Massachusetts 02139 \\ $^1$ current
address: University of Arizona, Tucson, Arizona 85721}
\date{\today}
\begin{abstract}

A new technique for maintaining high contrast in an atom
interferometer is used to measure large de Broglie wave phase
shifts. Dependence of an interaction induced phase on the atoms'
velocity is compensated by applying an engineered \emph{counter
phase}. The counter phase is equivalent to a rotation and
precisely determined by a frequency, and can be used to measure
phase shifts due to interactions of unknown strength. Phase shifts
of 150 radians (5 times larger than previously possible) have now
been measured in an atom beam interferometer and we suggest that
this technique can enable comparisons of atomic polarizability
with precision of one part in 10,000.
\end{abstract}
 \keywords{} \maketitle

Atom interferometers are now precision tools for measuring
interactions that cause a differential phase shift between atom
waves in two separated paths. For example, measurements of
platform rotation \cite{lhs97, gbk97} and acceleration
\cite{pcc99, pcc01}, gravity gradients \cite{mff02}, and atomic
polarizability \cite{esc95} have each been made using atom
interferometers to directly measure a corresponding atomic de
Broglie wave phase shift.  The precision of these measurements
made with cw atom beams can be improved by compensating for
dispersion. \emph{Dispersion}, i.e. a correlation between atom
wavelength and phase shift, has limited the interferometer in
\cite{lhs97,esc95} to a maximum of 35 radians of interaction
induced phase shift before contrast is reduced by $1/e$. In ref
\cite{gbk97} a rotation induced phase of 10 radians could reduce
the contrast by $1/e$. Here we present a new technique to maintain
high contrast while studying large de Broglie wave phase shifts
without reducing atom flux.  It consists of two separated regions
that induce time-dependant phases in a way that the net applied
phase depends on the atom's velocity.  We shall describe this
technique and show that it is equivalent to rotation at an
accurately known angular frequency.

    The source of contrast loss addressed here comes from
the experimental spread in atomic velocity combined with
dispersion. Most interactions are dispersive because the
interaction induced phase shift, or \emph{interaction phase},
depends on velocity to some power: $\phi_{int}(v) \propto v^n$.
The factor $n$ equals $-1$ for interactions with phase shifts that
depend on transit time, such as platform rotation or uniform
fields applied to one arm of an interferometer. The factor $n$
equals $-2$ for gravitationally induced phase shifts and also for
electric or magnetic field gradients across an interferometer made
with gratings. For $n \neq 0$, a spread in velocity leads to an
inhomogeneous phase and hence a loss of contrast.  Under such
circumstances the statistical power in a measurement of
interaction strength is optimized at a rather small interaction
phase.  For a Gaussian atomic velocity distribution, this phase
is: \mbox{$\phi_{int}(v_0) =
\huge{|}\frac{1}{n}\huge{|}\frac{v_0}{\sigma_v}$}, where $v_0$ is
the average and $\sigma_{v}$ is the r.m.s. width of the velocity
distribution. The supersonic atom beam in ref \cite{lhs97,pcc01}
has $\frac{v_0}{\sigma_v} = 25$, which limits the most sensitive
measurements to an interaction phase of 25 radians.

Dispersion compensation enables measurements of much larger
interaction induced phase shifts. We demonstrate this by using an
engineered  \emph{counter phase} to cancel dispersion, and regain
high contrast. The technique has many advantages for precision
measurements of an interaction strength---a very large interaction
phase can now be measured, all the atom flux is used, and the need
to precisely measure the velocity of the atoms is eliminated. It
is a quantum extension of the classical velocity multiplexing
technique that used mechanical choppers to modify the velocity
distribution of the beam \cite{hpc95}.

In an earlier proposal, Clauser \cite{cla88} noted that a magnetic
field gradient can compensate for gravitationally induced phase
shifts.  This idea in essence measures one interaction in terms of
another, the overall error being a combination of the two errors
separately. In contrast, the counter phase used here is determined
by a frequency that can be set to a precisely known and stable
value. Our technique is more closely related to methods developed
in \cite{kac92} and used in \cite{gbk97, pcc99, pcc01, mff02}
where dispersion compensation is achieved by moving the gratings
to simulate platform rotation or acceleration.

\begin{figure}
\epsfig{figure=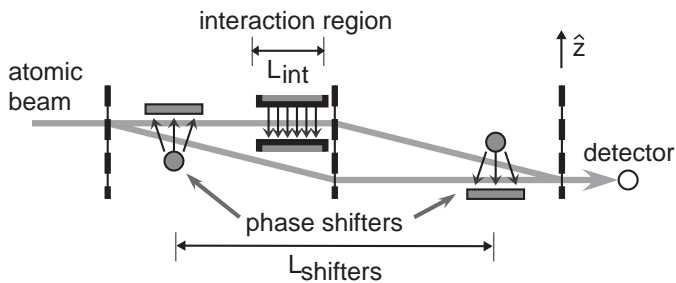, width=3.5in}
\caption{\label{fig:schematic} A schematic of the three-grating
atom interferometer. An interaction region is in the center, and
the two phase shifters used for dispersion compensation are
located on either side.}
\end{figure}

\begin{figure}
\epsfig{figure=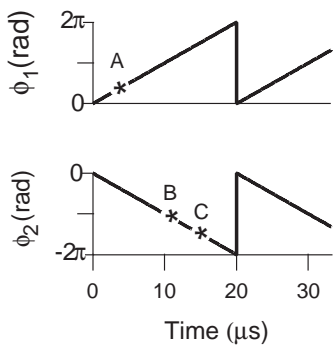, width=1.75in}
\caption{\label{fig:ramp}The time-dependent phases, $\phi_1(t)$
and $\phi_2(t)$, introduced by each phase shifter. The sum,
$\phi_1(t)+\phi_2(t)$ depends on velocity.  For example, faster
atoms may get phase shifts A and B, while slower atoms get the
phase shifts A and C.}
\end{figure}

To create the counter phase, two phase shift regions spaced a
distance $L_{shifters}$ apart are used to produce differential
phase shifts between the arms of the interferometer
(Fig.~\ref{fig:schematic}).  They apply a saw-tooth ramp ---
increasing the applied phase linearly from zero to $\pm2\pi$ and
then abruptly returning to zero---with frequency $f$. The phase
shifts must be opposite in sign, as shown in Figure \ref{fig:ramp}
and can be represented, modulo $2\pi$, as \be
\begin{split}
\phi_{1}(t) &= + 2\pi ft \\
\phi_{2}(t) &= - 2\pi ft.
\end{split} \label{eq:ramp} \ee
The sum of these phase shifts makes the counter phase (modulo
$2\pi$): \be
\begin{split}
\phi_{counter}(t,v) &= \phi_1(t)+\phi_2(t+L/v) \\
                    &= -2\pi f L_{shifters} /v.
\end{split} \label{eq:phicounter} \ee
Importantly the total counter phase depends on the atom's time of
flight between the two phase shifters, i.e. it has $v^{-1}$
dispersion, but is independent of time. The dispersion scales with
ramp frequency, hence it can compensate for a dispersive
interaction of any strength.  Fidelity of the two phase shifts
$\phi_{1,2}$ to Eq. \ref{eq:ramp} becomes less critical when the
ramp frequency is large as is discussed later.

The cancellation of dispersive effects in principle works
perfectly if the interaction phase is proportional to $v^{-1}$.
One type of interaction is a region of uniformly different
potential of length $L_{int}$ on one interferometer arm.  The
change in energy, $\hbar \omega_{int}$, for an atom inside the
field is the same for atoms of all velocities, but the phase shift
caused by the interaction depends on the transit time $L_{int}/v$
of the atom passing through the region: \be \phi_{int}(v) =
\omega_{int} L_{int} / v \label{eq:phiint} \ee With this
interaction alone, a spread in velocity from the atom source
creates a spread in $\phi_{int}$ that destroys the contrast of the
interference fringe and limits the size of $\omega_{int}$ that can
be measured.  By adding the counter phase, the total phase shift
is: \be \phi_{int}(v)+\phi_{counter}(v) = (\omega_{int} L_{int} -
2\pi f L_{shifters})/v \ee At the rephasing frequency, $f_{reph}
\equiv \omega_{int} L_{int}/2\pi L_{shifters},$ the total phase
shift is zero for all velocities.  There is no net dispersion, and
the fringe contrast should be ideal.

This method of rephasing the interference pattern can dramatically
improve a precision measurement of atomic polarizability such as
was made with the MIT interferometer \cite{esc95}. We can now
apply an interaction phase shift exceeding 150 radians for best
signal to noise, compared with 25 radians previously. Furthermore,
instead of measuring the atom beam velocity (and velocity
distribution) and then modelling the phase shift,  we need to
measure only the ramp frequency and the distance $L_{shifters}$.
In this proof-of-principle experiment, these advantages make it
possible to determine the polarizability of sodium atoms with the
same statistical precision as \cite{esc95} in $1/15$ the time.

\begin{figure}
\epsfig{figure=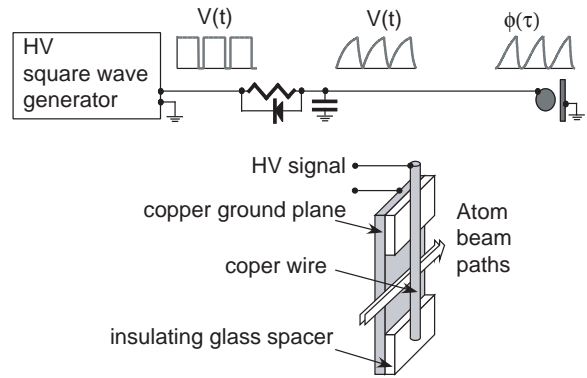,width=3.0in}
\caption{\label{fig:phaseshifter} A schematic of the electronics
driving each phase shifter, including a blowup showing how the
gradient field region is constructed.  Note that $\phi_1 \sim
V^2(t)$ is a good approximation to a linear sawtooth.}
\end{figure}

We have implemented the phase shifters by using small regions of
electric field gradient that create a phase shift by exposing each
path of the interferometer to a different field. The gradient
field is produced by a charged cylinder with radius $r=0.5$~mm at
a voltage of $V_0\approx 2$~kV. The paths of the interferometer
pass between the cylinder and a ground plane, which is a distance
$a=1.5$~mm from the cylinder axis (Fig.~\ref{fig:phaseshifter}).
The phase difference between the paths (separated by distance
$w\approx 50$~$\mu$m) is \be \phi_{1,2}(t) \approx \pm
\frac{\pi}{2} \; \ln^{-2}(\frac{2a}{r}) \; \frac{\alpha w}{v x^2}
V_0(t)^2
\ee 
where $x$ is the average distance of the two paths from the
cylinder axis and $\alpha$ is the polarizability of the atoms. The
two cylinders are oriented on opposite sides of the interferometer
so that they can apply opposite relative phases as required by
Eq.~\ref{eq:ramp}.  To create a linear ramp in phase, a voltage
must be applied to the cylinders with time dependence $V_0(t) \sim
\sqrt{t}$.  We approximate this ideal square-root shape in time by
filtering a rectangle-wave high voltage (with duty cycle $p
\approx 90\%$) using an $RC$ circuit. During the off-cycle a diode
drains the capacitor, quickly returning the voltage to zero. Using
this voltage waveform, the phase produced by the gradient fields
ramps approximately linearly from 0 to $2\pi$ during the on cycle:
\be \phi_{1,2}(t)=\pm \gamma (1-e^{-t/RC})^2 \qquad \text{for
}0<t<p/f. \label{eq:realramp} \ee The parameter $\gamma$,
proportional to the strength of the gradient field, can be changed
by adjusting the amplitude of the square wave or the position $x$
of each cylinder.  Setting to $\gamma=.83\pi$~rad and $RC=2.4/f$
best approximates the perfect ramp defined by Eq.~\ref{eq:ramp}.
High fidelity to Eq. \ref{eq:ramp} is not required if the ramp
frequency is much larger than the inverse time of flight between
the two phase shifters, since the deviation in phase is small
compared to the additional phase of $2\pi m$ corresponding to $m$
ramp cycles.

Applied to the measurement of polarizability \cite{esc95}, the
phase shifters are used in conjunction with an interaction phase
produced by a parallel plate capacitor (with voltage difference
$V$ and plate separation $d$) that makes a constant electric field
$\mathcal{E}$ surrounding one path of the interferometer for a
length $L_{int}$. The change in energy of an atom in the electric
field (with magnitude $\mathcal{E}=V/d$) is given by the
polarizability $\alpha$: \be \hbar \omega_{int} = \frac{1}{2}
\alpha V^2/d^2 \label{eq:omegaint} \ee

To determine $\alpha$, the phase-shifted interference pattern is
measured as oscillating atom beam intensity versus the transverse
position $z$ of one grating. For atoms of velocity $v$, the
interference pattern without a counter phase is \be I_v(z) = N + A
\cos [k_g z + \phi_{int}(v)], \ee where $N$ is the average
intensity, $A$ is the amplitude of the fringe, and
$k_g=2\pi/(\text{100 nm})$ is the grating wavevector.  The
measured interference pattern is a weighted average over the
velocity distribution  (which is approximately Gaussian with
$\sigma_v/v_0 \approx 0.04$, where $v_0$ is the average velocity
(1-2~km/s) and $\sigma_v$ is the rms velocity width): \be
\begin{split}
I(z)&=\int dv P(v) \big( N + A \cos [k_g z + \phi_{int}(v)] \big) \\
&=N + A C' \cos [k_g z + \phi'],
\end{split}
\label{eq:velocityaverage}
\ee
where the phase and contrast are
\be
\begin{split}
\phi' &= \phi_{int}(v_0)  \\
C'    &= \exp\Big{[}-\frac{1}{2}
\big{(}\frac{\sigma_v}{v_0}\big{)}^2 \phi_{int}^2(v_0)\Big{]},
\label{eq:dephasedcontrast}
\end{split}
\ee

\begin{figure}
\definecolor{ggray}{gray}{.5}
\definecolor{dgray}{gray}{.25}
\epsfig{figure=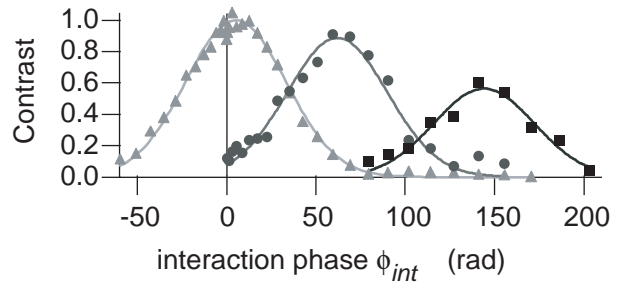,width=3.5in}
\caption{\label{fig:contrastplot} Contrast of the interference
pattern as a function of interaction induced phase,
$\phi_{int}(v_0)$. With no dispersion compensation
({\color{ggray}$\blacktriangle$}) contrast peaks near
\mbox{$\phi_{int}(v_0)=0$}. Using the time-dependent phase
shifters, contrast revives at \mbox{$\phi_{int}(v_0)=62$~rad} for
a ramp frequency $f=17$~kHz ({\color{dgray}$\bullet$}), and at
\mbox{$\phi_{int}(v_0)=144$~rad} for $f=40$~kHz ($\blacksquare$).}
\end{figure}

With the phase shifters and the interaction region both on, the
phase and contrast of the resulting interference pattern are: \be
\begin{split}
\phi' &= \phi_{int}(v_0) - 2\pi f L_{shifters}/v_0 \\
C'    &= \exp\Big{[}-\frac{1}{2}
\big{(}\frac{\sigma_v}{v_0}\big{)}^2 \big{(}\phi_{int}(v_0) - 2\pi
f L_{shifters}/v_0\big{)}^2\Big{]}. \label{eq:rephasedcontrast}
\end{split}
\ee Note this is merely a shifted Gaussian that has the same width
independent of ramp frequency. In principle the peak contrast
remains $C'=1$.

Measurements of contrast vs. interaction phase
(Fig.~\ref{fig:contrastplot}) were made with the phase shifters
ramping at fixed frequencies: $f =$ 0, 17.0, and $40.0$ kHz. A
revival in contrast occurs when $\phi_{int}(v_0) = 2\pi f
L_{shifters}/v_0$. However, the maximum rephased contrast is less
than the ideal 100\% because of imperfections in the phase
shifters such as non-linearity of the implemented phase ramp
Eq.~\ref{eq:realramp}. The rightmost curve ($\blacksquare$) in
Fig.~\ref{fig:contrastplot} demonstrates that polarizability can
now be measured at an interaction phase as large as 150~rad.

\begin{figure}
\epsfig{figure=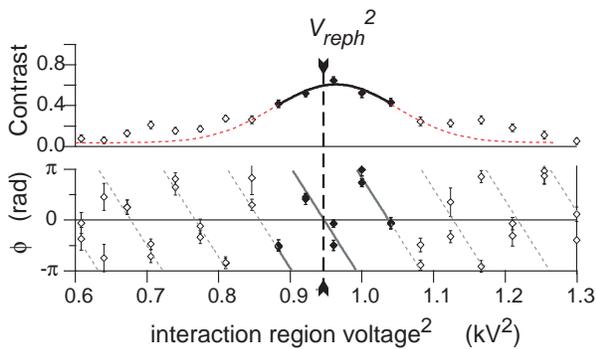, width=3.5in}
\caption{\label{fig:phaseplot} The relative contrast and phase as
a function of the voltage-squared in the interaction region.
Atomic polarizability can be determined by finding the voltage,
$V_{reph}^2$, at which the fitted total phase is zero (and the
contrast is maximum). The fit was made only to the 10 central data
points ($\blacklozenge$). The dashed line is an extrapolation from
the fit, shown for comparison with the data outside the fit region
($\lozenge$).}
\end{figure}

Polarizability can be determined using this technique by finding
the parameters at which the total phase is exactly zero. Then,
there is high contrast, and no explicit dependance on velocity:
\be \omega_{int} L_{int}/v = 2\pi f L_{shifters}/v.
\label{eq:rephasecondition} \ee Phase measurements are shown in
Fig.~\ref{fig:phaseplot} and are fit to a line whose zero crossing
determines $V^2_{reph}$. Then the polarizability is: \be \alpha =
\frac{2 h f L_{shifters} d^2}{L_{int} V_{reph}^2}.
\label{eq:newalpha} \ee

The precision of the polarizability measurement depends on the
precision $\Delta\phi'$ with which the interaction phase can be
measured, and also on $\phi_{int}(v_0)$: \be
\frac{\Delta\alpha}{\alpha} = \frac{\Delta\phi'}{\phi_{int}(v_0).}
\label{eq:alphaerror} \ee While the statistical uncertainty in
phase grows as the contrast is lost, $\Delta\phi' \sim 1/C'$, the
fractional error in polarizability is reduced by making the
interaction phase $\phi_{int}(v_0)$ larger.

\definecolor{ggray}{gray}{.5}
Without the counter phase (e.g. in the leftmost curve
({\color{ggray}$\blacktriangle$}) of Fig.~\ref{fig:contrastplot})
the best signal-to-noise ratio occurs at an interaction phase of
$\phi_{int}(v_0)=v_0/\sigma_v=25$~radians. Precision in
$\Delta\phi'$ here is $\sim$800~mrad$/\sqrt{\text{sec}}$, which
makes us require roughly 300~sec of measurement time to achieve a
fractional precision in polarizability of
$\Delta\alpha/\alpha=0.2\%$.

With the counter phase we achieve the same 0.2\% statistical
precision in $\alpha$ with only 20~sec of measurement. The ten
phase measurements used in the linear fit \mbox{($\blacklozenge$
data points in Fig.~\ref{fig:phaseplot}} constitute a total of
20~sec of measurement, and the resulting uncertainty in the
interaction phase is $\Delta \phi' = 130$~mrad, while the
interaction phase itself is $\phi_{int}(v_0)=66$~rad. Thus, the
fractional uncertainty in polarizability is $\frac{\Delta
\alpha}{\alpha} = \frac{130~mrad}{66~rad} = 0.2\%$. Using the
contrast revival demonstrated at $150$~rad, it should be possible
to measure the interaction phase to $10^{-4}$ statistical
precision with 50 minutes of data.

Sources of systematic error include uncertainty in $L_{shifters}$,
the calibration of the phase shifts $\phi_{1,2}$, and the
dimensions of the interaction region.  (The frequency $f$ can be
determined essentially without errors.) The distance
$L_{shifters}$ is on the order of $1$~m and can be measured to one
part in $10^{4}$ with a ruler, or much better if measured
interferometrically. The most difficult new systematic errors to
analyze stem from deviations in the ramped phase from the ideal
form in Eq.~\ref{eq:ramp}.  If the ramp maxima deviate from $\pm
2\pi$ by $\epsilon$ but are at least symmetric, i.e. $\phi_1(t) =
- \phi_2(t)$, the corresponding error in $\alpha$ is smaller than
$\epsilon/2 \pi$ by a factor equal to $L_{shifters} v^{-1}f^{-1}$.
This is because each time $\phi_2(t)$ returns to zero during an
atom's time of flight, it is equivalent to adding a precise
integer times $-2 \pi $ to the counter phase. However, if
$\phi_1(t) \neq - \phi_2(t)$ the error in phase will be: \be
\phi_{error} = \langle \phi_1(t)+\phi_2(t) \rangle_t \ee where
$\langle \ldots \rangle_t$ represents an average over one period
of the ramp signal \cite{rob02}.  To correct for this,
$\phi_{error}$ was measured in steady state, i.e. with $f=0$, and
both phase shifters on.

The dispersion compensation technique presented here eliminates
two of the three major sources of uncertainty that limited the
precision in $\alpha$ obtained in \cite{esc95} (i.e. contrast loss
and uncertainty in the velocity distribution).  However a
measurement of sodium's polarizability at the $10^{-4}$ level will
still be difficult due to uncertainty in the geometry of the
interaction region. To attain $10^{-4}$ precision in the parameter
$d^2/L_{int}$ in Eq.~\ref{eq:newalpha}, the dimensions must be
measured to within $10^{-7}$ meters and the electric field
modelled very accurately. New techniques can meet these
requirements \cite{rob02}.  However, by measuring the the
\textit{ratio} of polarizabily of two alkali atoms using the same
interaction region and phase shifters, uncertainties in all the
geometrical lengths ($L_{shifters}$, $L_{int}$, and $d$) will be
common to both so the resulting ratio of polarizabilities will be
limited only by statistical errors if the atoms traverse the same
path through the interaction region.

Used as a Sagnac gyroscope the interferometer is also sensitive to
rotations which create a $1/v$-dependent phase shift: \be
\phi_{int}(v) = 2 k_g L_g^2 \Omega /v \ee where $\Omega$ is the
rotation rate and $L_g$ is the distance between the
interferometer's gratings, and $k_g$ is the wave number of the
gratings \cite{lhs97,gbk97}. Our technique exactly cancels such
dispersion and may therefore be regarded as equivalent to
physically rotating the interferometer.  Thus, one could servo the
engineered phase and use the integral of the rephasing frequency
as a measurement of accumulated angular displacement.

The rephasing technique will also work to some extent for
interactions with \textit{any} velocity dependence, $\phi_{int}(v)
\propto v^n$.  For example, the interaction phase with optimum
signal to noise can be increased by the factor $ v_0/|n+1|\sigma_v$ 
since \be \phi_{int}(v_0) <
|\frac{1}{n(n+1)}|(\frac{v_0}{\sigma_v})^2 \ee is allowed by using
the $v^{-1}$ counter phase presented here.

In conclusion, we have demonstrated a new method to maintain high
contrast of the interference pattern while studying large de
Broglie wave phase shifts in an atom interferometer. The method
will be of immediate benefit to new precision measurements of
atomic polarizability. In addition, the technique should serve to
restore contrast in any experiment where the path length
difference grows larger than the atomic coherence length.

\bibliography{DC}
\end{document}